\documentstyle[preprint,aps,epsfig]{revtex}
\global\firstfigfalse

\begin{document}

\draft
\preprint{AZPH-TH/96-16}

\title{What is the Brightest Source for Dilepton Emissions at RHIC?}

\author{David Fein, Zheng Huang\footnote{Electronic address: 
huang@physics.arizona.edu},
Peter Valerio and Ina Sarcevic}

\address{Department of Physics,
University of Arizona, Tucson, AZ 85721, USA}

\date{July, 1996}
\maketitle
\begin{abstract}
We calculate the dilepton emissions as the decay product of
the charm and bottom quarks produced in heavy-ion collisions 
at RHIC energy. We take into account the next-to-leading-order 
radiative corrections in perturbative QCD to the heavy quark 
production from both an initial hard parton-parton scattering 
and an ideal quark-gluon plasma. We find that the thermal charm 
decay dominates the dilepton production in the low dilepton 
mass region ($<2$ GeV), while the heavy quark production from the 
initial scattering takes over the intermediate and high mass 
regions ($> 2$ GeV). Our result also indicates the importance 
of the bottom quark in the high mass region ($>4 $ GeV ) due to 
its large mass and cascade decay. If the initial scattering 
produced charm suffers a significant energy loss due to the 
secondary interaction, the bottom decay constitutes the major 
background for the thermal dileptons.
\end{abstract}

\newpage
The emission of dileptons in  high energy heavy-ion collisions
provides an excellent probe of the property of  dense hadronic
matter such as the quark-gluon plasma and the hot hadronic gas. Due to
the small cross sections of electromagnetic interactions, these
dileptons once produced can escape the strong interacting volume
of  sizes which might be produced in heavy-ion collisions. The
dilepton invariant mass spectrum 
holds the most promising signatures of the quark-gluon plasma
such as  thermal dileptons \cite{shuryak} and the J/$\psi$
suppression \cite{satz}. However, the use of the dilepton probe
is difficult for an obvious reason: it is sensitive to many different
sources. In the low mass region, the resonance decays from the light
hadrons constitute the main background. In the intermediate and high
mass regions, the Drell-Yan dilepton production from the initial 
hard scattering is important. In high energy heavy-ion colliders such
as RHIC and LHC, the heavy flavor quark production can be quite substantial
and its subsequent decay can lead to a large combinatorial background
for the dilepton spectrum. Therefore it is crucial to make a solid theoretical
prediction on the contribution from  heavy quark decays. 

In the Letter we discuss the dimuon
 emission as the 
decay product of the heavy
quarks (charm and bottom) produced in an initial hard scattering
and in a thermalized quark-gluon plasma at  RHIC energy.
We include the next-to-leading 
order radiative corrections and nuclear shadowing effects. 
Indeed, the recent work by Gavin, McGaughey, Russkanen and Vogt
\cite{gavin} has 
indicated that the dilepton production from the correlated charm decay
is by far the dominant contribution. However, if the  
energetic charmed parton is subject to energy losses due to multiple
secondary interactions in the dense matter produced in the collisions,
the dilepton spectrum may be considerably softened, as suggested
recently by Shuryak \cite{shur2}. In contrast to the charm, the bottom
decay can yield higher invariant mass lepton pairs even at rest due
to the large bottom mass.
It is thus
necessary to examine the various components of the heavy quark spectra and
their role in the dilepton production in an extended dilepton mass
region.
We also examine the sensitivities of the perturbative calculation
on the choices of the relevant physical scale
and the heavy quark fragmentation function. The heavy parton production
is calculated using the explicit next-to-leading order 
matrix elements, while the
parton fragmentation and the heavy hadron decay are handled by
the {\sc Isajet} Monte Carlo model \cite{isajet}.
The cascade decay of the bottom quark, $b\rightarrow c\rightarrow s$
is also included in this model.

Theoretical calculations of heavy quark differential and total cross
sections to the next-to-leading order $O(\alpha_s^3)$
at RHIC and LHC energies have been performed by several groups \cite{hq}
for the initial hard scatterings. It is found that higher order radiative
corrections are important and the theoretical $K$-factor
ranging from $1\sim 3$ has strong transverse momentum dependence. The 
nuclear shadowing effect is also non-negligible even at RHIC energy.
Therefore, it is crucial to include full next-to-leading order 
corrections when one evaluates the dilepton production from heavy-quark
decay. 

The production of heavy hadrons is based on the factorization 
theorem, which states that the cross section for the production
process of a hadron with an energy-momentum $(E,{\bf k})$ can be written in
a factorized form
\begin{equation}
E\frac{d\sigma}{d^3k}=\sum_i\int E'\frac{d\sigma_i}{d^3k'}(k',\mu )
D_i^{(h)}(z,\mu )\frac{E}{E'}\frac{dz}{z^3}\; ,\label{1}
\end{equation}
where $k'=k/z$, $D_i^{(h)}(z,\mu )$ is the fragmentation function
for producing the hadron $h$ from the parton $i$ and $d\sigma_i$ is
the short-distance cross section for producing the parton $i$ from
the colliding hadrons. Clearly
if $i$ is a light parton, the cross section $d\sigma_i$ is not finite
and the  prescription for subtracting the collinear singularity has to be
employed. However, in the case of heavy quark production,
$d\sigma_i$ is an infrared
safe quantity because the heavy quark mass acts as a cutoff for final
state collinear singularities. Thus, the heavy quark mass sets the scale
for a perturbative expansion in the strong coupling constant. In the 
limit when the mass can be considered much larger than the typical
hadronic scale, the fragmentation function $D_i^{(h)}(z,\mu )$ is
also calculable in perturbation theory. As long as the transverse
momentum of the produced heavy quark is not much larger than its mass,
perturbation theory predicts $D_i^{(h)}(z)\sim \delta (1-z)$. On the
other hand, the finite mass and energy transfer ($\Delta E$)
in the fragmentation
process suggests that the breakup amplitude $\sim$ $1/\Delta E$, where
\newpage
\begin{eqnarray}
\Delta E & = &(m_Q^2+z^2k^2)^{1/2}+(m_q^2+(1-z)^2k^2)^{1/2}
  -(m_Q^2+k^2)^{1/2}\nonumber \\
& \propto & 1-1/z-\epsilon_Q/(1-z)\; .
\end{eqnarray}
This leads 
to a Peterson form \cite{peterson}
\begin{equation}
D_Q^{(H)}=\frac{N}{z[1-(1/z)-\epsilon_Q/(1-z)]^2}\; ,\label{2}
\end{equation}
where $\epsilon_c=0.15$ and $\epsilon_b=0.016$. The normalization
constant $N$ is fixed by the sum rule $\int dz D_Q^H (z) =1$:
$N=1.28$, 0.25 for $c$ and $b$ respectively. 
Such a parameterization has been shown to be appropriate in
$e^+e^-$ reactions and fit well the MARK-II data \cite{peterson}. For
hadronic collisions or an even heavier quark decay, it is known that
there are no universal fragmentation functions. 
The double differential
cross sections for (all) charmed hadrons (mainly $D$-mesons) 
as well as (all) 
$b$-hadrons (mainly $B$-mesons) are calculated using
Eq.(\ref{1}) (assuming that there is no energy loss for the
produced parton before hadronization) 
and the results are plotted in Fig.1 against
the distributions for the heavy partons. 

The initial parton differential
distributions are  calculated using the next-to-leading-order (NLO)
matrix elements provided by Nason, Dawson and Ellis in \cite{nlo}.
We have taken $m_c=1.5$ GeV, $m_b=4.7$ GeV, the factorization
and renormalization scale $\mu$ to be the transverse mass $2m_t$. 
The nucleon structure function is parameterized by MRS A and
the nuclear shadowing parametrization 
fits the EMC, NMC and E665 data \cite{qiu}.
The total cross sections (per initial nucleon) 
are $\sigma_{c\bar{c}}^{\rm tot}=153\mu$b
and $\sigma_{b\bar{b}}^{\rm tot}=1.5\mu$b for Au+Au collisions at
$\sqrt{s}=200$ GeV/n. The choice of a smaller charm quark mass
$m_c=1.2$ GeV increases the total cross section by a factor of 2
(see Vogt in \cite{hq}).
\begin{figure}
\centerline{\epsfig{figure=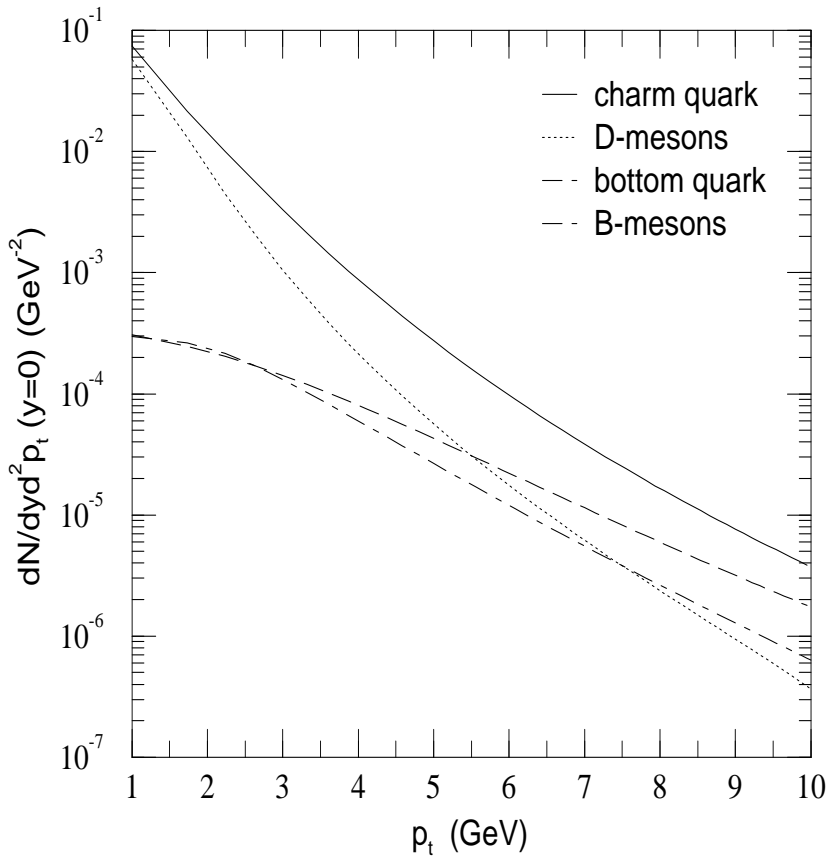,width=5.5in,height=5in}}
\caption{The heavy quark hadron differential cross sections
calculated using the Peterson heavy quark fragmentation
functions convoluted with  the parton cross sections.
The initial heavy quark productions are calculated using the
explicit next-to-leading-order matrix elements with MRS A structure
function and the nuclear shadowing function parameterized by
Benesh {\it et. al.} [9]}
\end{figure}
As is evident from Fig.1, 
 the leading
order (in heavy quark mass) fragmentation function  
$D(z)\sim \delta (1-z)$ seems to be a reasonable approximation for 
heavy quarks.
There is some noticeable softening of heavy mesons at large $p_t$
and an enhancement of $B$-meson at very small $p_t$ due to the 
finite heavy quark mass and the heavy quark-hadron mass splitting.  

The semileptonic decay of the heavy meson can be  conveniently
handled by  Monte Carlo simulations such as {\sc Isajet} 
or {\sc Jetset}. These Monte Carlo programs usually 
have the leading order matrix elements as the default value while 
the initial and final state radiations are treated by a parton
shower model. The virtue of the parton shower algorithm is that
it is beyond the leading logarithmic approximation and incorporates
the multiple parton emissions. On the other hand, it is not a consistent
higher order calculation since it does not contain the virtual
corrections in the matrix elements. Simply embedding the NLO
 matrix elements in the Monte Carlo program may cause an
ambiguity due to double counting of multiple parton emissions.
To circumvent this difficulty and 
to preserve the NLO
heavy quark production calculation, we develop a numerical routine
to generate the events at the parton level. We explicitly integrate
out the initial parton momenta in the total cross section 
Monte Carlo integral and write out the four-momentum for each
heavy quark. 
These momenta are associated with the corresponding
weight in the phase space multiplied by the corresponding
value of the transition probability. Summing up these
weights gives the total cross section. 
One can then use  {\sc Isajet} to
fragment each parton pair and to track the subsequent 
correlated decays.  

When generating events, one encounters a genuine feature of 
the higher-order calculation: the presence of the terms needed to
cancel the collinear singularities in the initial state radiation.
In the factorization scheme, one uses
the ``+ function'' prescription to calculate these terms.
As a result, the events associated with the substraction terms have
to be generated in the Monte Carlo integration of the matrix 
elements.
There is no guarantee that the 
differential cross section should always add up to a positive 
value everywhere in the presence of the substraction terms. 
Some  negative bins may appear in the region close to the boundary
of phase space for real emission. The problem is worse when one
calculates the dilepton spectrum.  Due to the statistical nature
of the decay process, there is no prediction where the negative 
bins may appear in the dilepton invariant mass spectrum. In order
to make the distribution more readable in the presence of 
such negative bins, we will always enlarge the size of the these bins
so as to integrate over more and more positive real contributions.

At sufficiently high temperatures, a thermal system of light
partons can create a charm pair which will  decay into leptons
and should also be included. 
The thermal charm production has been calculated so far at the leading order
matrix element level \cite{muller}. To estimate the reliability of the
leading-order result, one has to perform a full NLO calculation. 
At the next-to-leading order, there are
additional $2\rightarrow 3$ processes and loop diagrams. Formally,
the differential production rate is given by the convolution of the
parton level differential cross sections with the initial light parton
phase space densities (distribution functions) $f_i(p)$ (we neglect
the final state Fermi blocking or Bose-Einstein enhancement\footnote{
When the final state quantum statistical effect is present, one also needs
to consider $3\rightarrow 2$ processes in order to organize the 
collinear singularity in a familiar factorizable form. We do
not consider this possibility.} )
\begin{eqnarray}
E\frac{dR}{d^3k} & = & \int \frac{d^3p_1}{(2\pi )^32E_1}
\frac{d^3p_2}{(2\pi )^32E_2} F(p_1,p_1) \left[ 
\frac{g_{gg}}{2}f_g(p_1)f_g(p_2)E\frac{d\hat{\sigma}_{gg}}{d^3k} +\right.
\nonumber \\ 
 & & \left. g_{q\bar{q}}f_q(p_1)f_{\bar{q}}(p_2)
E\frac{d\hat{\sigma}_{q\bar{q}}}{d^3k}+
g_{qg}f_q(p_1)f_g(p_2)E\frac{d\hat{\sigma}_{qg}}{d^3k}\right] \; , 
\label{rate} 
\end{eqnarray} 
where $g_{gg}=16^2$, $g_{q\bar{q}}=6^2N_f$ and $g_{qg}=6\cdot 16N_f$
are the degeneracy factors, and $F(p_1,p_2)=4p_1p_2$ is the 
inverse flux factor.

At the leading order, the $d\hat{\sigma}_{qg}$ term is absent. At the 
next-to-leading order, 
these parton level differential cross sections contain $2\rightarrow 3$
processes and loop diagrams and they 
are not infrared finite quantities. Braaten and Pisarski \cite{pisarski}
have shown that a resummation technique has to be employed to take
into account the so-called ``hard thermal loops'' which give the same
order of magnitude contributions as the leading order terms
when the external momenta
of the amplitude are soft. As a result of the resummation, the infrared
singularities are shielded by a Debye screening effect,  and the
screening masses of gluons and quarks provide an infrared cutoff. 
The perturbative resummation can only regularize the electric
part of the propagators and so far the magnetic screening mass has to 
come from the unknown nonperturbative effect. To obtain
 a qualitative result for the NLO thermal charm production,
we use the Debye screening mass \cite{wang}
\begin{equation}
\mu_D^2=\frac{6g^2}{\pi^2}\int pf(p)dp=4\pi\alpha_sT^2\; , \label{debye}
\end{equation}
to regularize all singularities in the radiative cross sections. This
amounts to a ``massive gluon scheme'' where the $\mu_D^2$ sets the
scale for the parton level cross sections in Eq.(\ref{rate}). 
Certainly, this calculation is not entirely consistent, and the results
thus obtained can only be taken as an qualitative indication 
on the reliability
of the leading-order result.
\begin{figure}
\centerline{\epsfig{figure=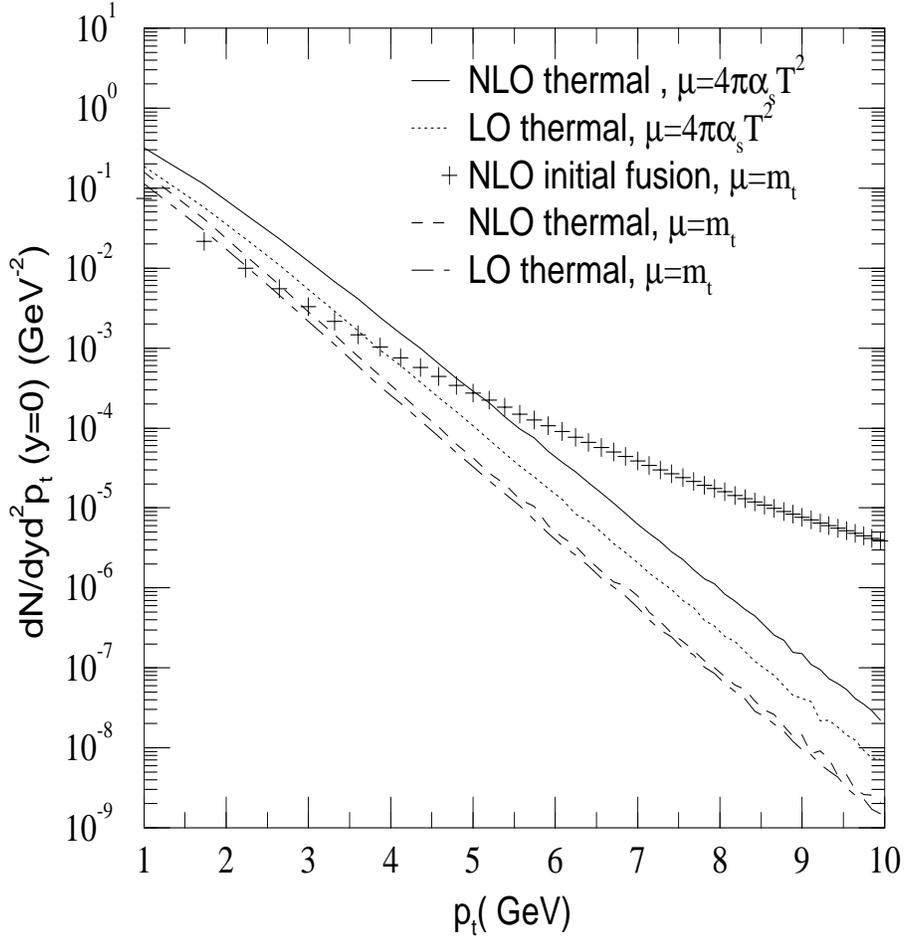,width=6in,height=5.5in}}
\caption{The charm differential distribution produced in an ideal
thermalized quark-gluon plasma at the leading and the next-to-leading 
orders in perturbative QCD. The sensitivity on the choice of the
scale is illustrated. The charm production at the NLO from the initial
hard scattering is also plotted for comparison.} 
\end{figure}
We have evaluated the thermal charm yield at RHIC energy by integrating
over the space-time evolution of the quark-gluon plasma. We assume 
a fully equilibrated thermal system with an initial temperature 
$T_0=0.55$ GeV and a Bjorken ideal-fluid scaling 
$T(\tau )/T_0=(\tau_0/\tau )^{1/3}$ with $\tau_0=0.7$ fm/c and $T_c=0.2$ GeV.
A spatial rapidity cutoff $\eta_{\rm max}=3$ is assumed. The differential
distributions are plotted in Fig.2 for the leading order and the
next-to-leading order calculations using
the Debye screening scale in (\ref{debye}). 
We also compare the result for a choice of the scale $\mu =m_t$.
As in the initial hard scattering
case, the higher order corrections are sizable with an effective $K$-factor
ranging from $1.7$ to $2.5$ as $p_t$ varies from
$1$ GeV to $10$ GeV. We use the leading order kinematics to generate the
heavy quark momenta and multiply the weights uniformly by an average
value $K=2$ to account
for the radiative corrections. 

The calculated dimuon spectra from the decay of heavy quarks
produced in a hard scattering and in an idealized quark-gluon plasma
are plotted in Fig.3. 
The dimuon spectrum from thermal charm quark production is considerably
softer than the initial hard scattering results 
as can be inferred from the $p_t$ distribution in Fig.2. However, it
dominates the dilepton production in the low mass region ($M<2$ GeV).
This can be a important background at RHIC for the dilepton 
signature for the low lying vector resonances such as $\rho$ and $\phi$.
It is important to note that the calculated thermal
charm contribution corresponds to a very ideal case where the 
quarks and gluons are fully equilibrated. The result may be considerably
smaller in the plasma where the chemical process is not in equilibrium
\cite{muller}.  
The thermal production of the bottom quark is negligible and is not
included in Fig.3.
The cascade decay of $B$-mesons, $b\rightarrow c\rightarrow s$
is also included in the curve for the initial fusion bottom production.
We have not assumed any secondary interactions in the medium produced
in the collision.
Although the production rate for the initial bottom quark is much smaller
than the charm, the dilepton production from $B$-meson decay is
non-negligible in the 
large invariant mass region due to the large
rest mass of the $B$-meson\footnote{ 
The results in \cite{gavin} give a larger
initial charm contribution due to the choice of a
smaller charm mass. However, we do
not have a complete understanding of their  
rather small bottom contribution.}.  
It becomes the dominant contribution when the dimuon invariant mass
is above $\sim 4$ GeV.
If the fast charm suffers the significant energy loss before it
hadronizes, the dimuon spectrum from the initial fusion charm
can be considerably softened.
In fact, it becomes very similar to that of the thermal charm
and falls off rapidily as $M$ gets bigger than 2 GeV, 
as demonstrated by Shuryak \cite{shur2}.
In this case, the bottom dominance is more prominent and starts
at a smaller value of $M\sim 2$ GeV because two $B$-mesons at
rest can produce an energetic lepton pair. 
 To make a comparison, 
 the Drell-Yan dimuon production taken from 
\cite{gavin} is also indicated in the plot.
\begin{figure}
\centerline{\epsfig{figure=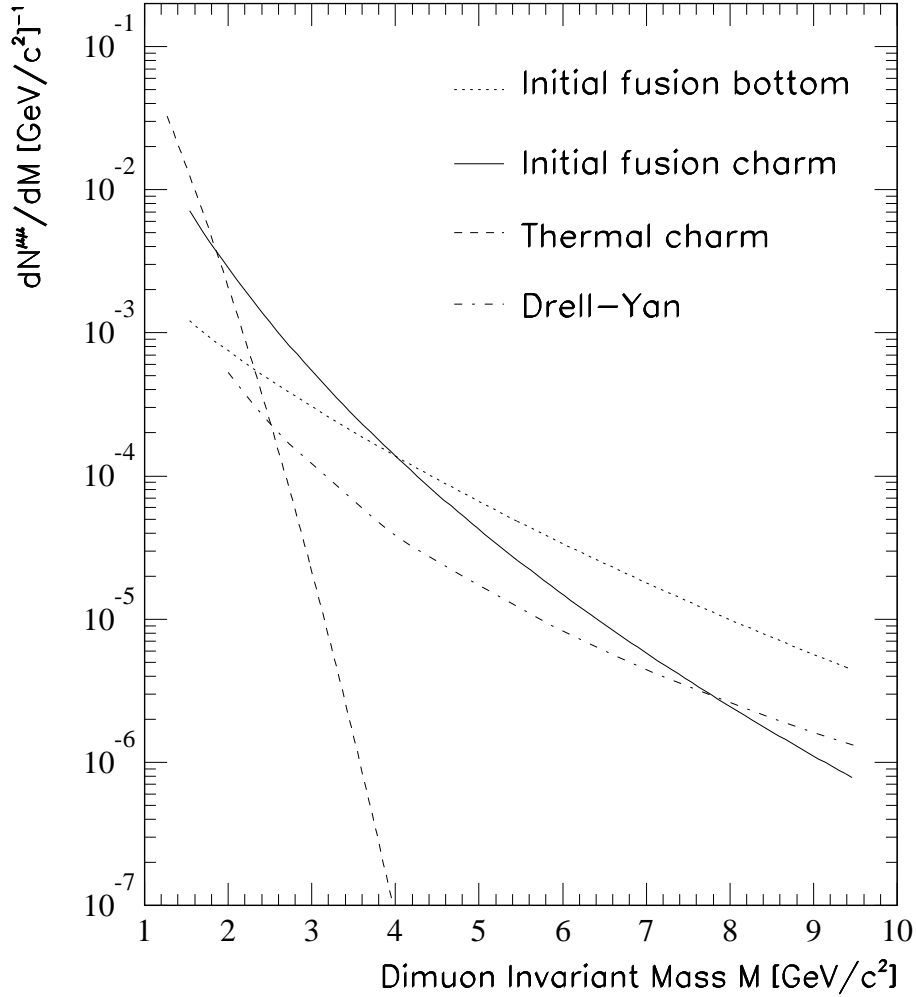,width=14cm,height=16cm}}
\caption{The dimuon invariant mass spectra
from the correlated heavy quark decay calculated using {\sc Isajet}.
No energy loss for the fast parton is assumed.}
\end{figure}
In conclusion, we have evaluated the dilepton emissions from
the decay of the heavy quark productions from the initial
hard scatterings and from a quark-gluon plasma in Au+Au collisions
at RHIC energy $\sqrt{s}=200$ GeV/n. 
We check the reliability of the leading order results
by explicitly performing a next-to-leading-order calculation.
The thermal charm from an
ideal quark-gluon plasma dominates the dilepton emission in the
low dilepton invariant mass region ($<2$ GeV). The correlated
heavy mesons 
produced  from
the initial hard scattering constitutes the major contribution to
the intermediate and high mass dileptons. If there is significant
energy loss for the fast charm, the bottom decay is by far the
dominant component in the intermediate and high dimuon mass regions.
In any case, we are likely to find more bottom than charm in the
high mass dilepton channel in central Au+Au collisions at RHIC. 

We would like to thank R.\ Vogt for useful discussions.
Z.H.\ and I.S.\ Thank the Aspen Center for Physics for
hospitality where part of this work is
accomplished. 
This work was supported in part by 
the U.S. Department of Energy Contract No.\ 
DE-FG03-93ER40792.
 
\end{document}